\documentstyle[12pt,aas2pp4,epsfig]{article}

\def\be{\begin{equation}}
\def\ee{\end{equation}}
\def\ea{{\it et al.}\,}
\def\eg{{\it e.g.},\,}

\def\apj{{\it ApJ}\,}

\def\rel{relativistic \,}

\begin{document}

\hfill\today
\title{Results from a Second RXTE Observation of the Coma Cluster}

\author{Yoel Rephaeli\altaffilmark{1,2}, and Duane Gruber\altaffilmark{3}}

\affil{$^1$Center for Astrophysics and Space Sciences, 
University  of California, San Diego,  La Jolla, CA\,92093-0424}

\affil{$^2$School of Physics and Astronomy, 
Tel Aviv University, Tel Aviv, 69978, Israel}

\affil{$^3$4789 Panorama Drive, San Diego CA 92116}

\begin{abstract}
The RXTE satellite observed the Coma cluster for $\sim$177\,ks 
during November and December 2000, a second observation motivated by 
the intriguing results from the first $\sim$87\,ks observation in 
1996. Analysis of the new dataset confirms that thermal emission from 
isothermal gas does not provide a good fit to the spectral distribution 
of the emission from the inner 1$^o$ radial region. While the observed 
spectrum may be fit by emission from gas with a substantial temperature 
gradient, it is more likely that the emission includes also a secondary 
non-thermal component. If so, non-thermal emission comprises $\sim 8\%$ 
of the total 4--20 keV flux. Interpreting this emission as due to Compton 
scattering of relativistic electrons (which produce the known extended 
radio emission) by the cosmic microwave background radiation, we determine 
that the {\it mean, volume-averaged} magnetic field in the central region 
of Coma is $B \sim 0.1-0.3$ $\mu G$.
\end{abstract}


\section{Introduction} 

X-ray spectra of clusters of galaxies have long been expected to show 
structure beyond that of a single temperature thermal model, mainly due 
to non-isothermality of intracluster (IC) gas in the outer cluster 
region. In addition, non-thermal (NT) X-ray emission in clusters was 
predicted (\eg, Rephaeli 1977) from Compton scattering of relativistic 
electrons by the Cosmic Microwave Background (CMB) radiation. There is 
at least some observational evidence for radial variation of the gas 
temperature in a few clusters (\eg, Markevitch 1996, Honda \ea 1996, 
Donnelly \ea 1999, Watanabe \ea 1999). In the Coma cluster, recent XMM 
measurements indicate (Arnaud \ea 2001) that the temperature is 
remarkably constant within the central region where temperature 
variation was previously deduced from ASCA measurements. After a long 
search (for a recent review, see Rephaeli 2001), NT X-ray emission seems 
to have finally been measured in Coma (Rephaeli, Gruber \& Blanco 1999, 
hereafter RGB, Fusco-Femiano \ea 1999), A2256 (Fusco-Femiano 2000), 
A2319 (Gruber \& Rephaeli 2002), and perhaps also in A2199 (Kaastra 
\ea 2000).

Appreciable deviation from isothermality may have significant impact on 
modeling the structure and evolution of IC gas, and on use of the gas 
as a probe to determine the total cluster mass (assuming hydrostatic 
equilibrium). The exact gas density and temperature profiles are also 
very much needed in analysis of measurements of the Sunyaev-Zeldovich 
(S-Z) effect and its use as a cosmological probe. There clearly is 
strong motivation for a more realistic characterization of cluster 
X-ray spectra for an improved description of IC gas, and the study of 
NT phenomena in clusters. 

We have previously analyzed $\sim 87$ and $\sim 160$ ks RXTE 
measurements of the Coma cluster and A2319, respectively, in order to 
search for NT emission from these clusters which have well documented 
extended regions of radio emission. Analyses of these measurements 
yielded strong evidence for a second spectral component in both 
clusters. While the second component could possibly indicate a 
temperature variation across the cluster, we have concluded (RGB, 
Gruber \& Rephaeli 2002) that the deduced spectral parameters are more 
consistent with power-law emission. NT emission in Coma seems to 
have been detected directly -- in the 25-80 keV range -- by BeppoSAX 
(Fusco-Femiano \ea 1999). This provided further impetus to propose a 
longer observation of this cluster with RXTE. Here we briefly report 
the results from a joint analysis of these and the previous RXTE 
measurements, with a total integration time of $\sim 264$ ks, and 
discuss some of their direct implications.

\section{Observations and Data Reduction}

Coma was observed with the Proportional Counter Array (PCA) and the High 
Energy X-ray Timing Experiment (HEXTE) on RXTE during 58 separate 
pointings totaling approximately 87 ks in June 10 -- 22, and July 15, 
1996, and during 50 additional pointings totaling nearly 177 ks in 
November 24 -- December 15, 2000. Spectral results from the earlier 
observation were reported by RGB.

PCA data were collected in the `Standard 2' spectral mode, which 
consists of a 129-channel count spectrum nominally spanning 2 to 
1000\,keV with 16 second time resolution. Data from the two independent 
HEXTE clusters of detectors were taken in event-by-event mode, which 
were subsequently accumulated into 256-channel spectra spanning 
17--250\,keV. To subtract the background, each HEXTE cluster was 
commanded to beamswitch every 16\,s or 32\,s between on-source and 
alternate off-source positions 1.5$^{\circ}$ on either side. 

Standard screening criteria were applied to the data segments (Earth
elevation angle, spacecraft pointing, avoidance of the South Atlantic
Anomaly, times of geomagnetic activity), resulting in a net exposure 
time for PCA of 87328\,s during the 1996 observations and 176864\,s 
during the second observation set. During the first set of observations 
PCA detectors 0, 1 and 2 were always on, but detectors 3 and 4 were 
enabled for less than half of the observation set, and were ignored in 
the interest of a more uniform data set. The (year) 2000 observations 
were made with PCA detectors 0 and 2 only, but the detector 0 spectra 
could not be used because of the loss of the guard propane layer, 
resulting in higher and uncertain background. The detector 2 gain 
was about 10\% higher in the later observations. The HEXTE data, by 
contrast, formed a perfectly uniform data set, with no changes of 
number of detectors or gain. HEXTE net times were shorter than PCA by 
nearly a factor of four: half the time was used for background 
measurement, and nearly another factor of two was lost due to 
electronic dead time caused by cosmic rays.

The PCA background was estimated with the `L7/240' faint source model
provided by the instrument team. No significant emission from Coma was 
detected above $\sim 40-50$ keV. The spectral data above 40 keV were 
used to determine a correction of $\sim$0.5\% to the background 
estimate. No such tuning was required for the HEXTE background, which 
is measured, and has been determined (MacDonald 2000) to be accurate 
to within a few hundredths of a count/s in long exposures.

\section{Spectral Analysis}

At about 40 count/s per PCA detector, the observed counting rate from
Coma was well above the background rate of 13 count/s per detector, thus 
the cluster was easily detected and possible errors of background 
estimation are unlikely to compromise spectral analysis. The HEXTE
net rate of somewhat less than one count/s per cluster was well below 
the background rate of 80 counts/s per cluster but still large compared
to the error limit of roughly 0.03 counts/s determined by MacDonald 
(2000). Inspection of the screened light curves for PCA and HEXTE 
revealed no significant variation, as expected for a cluster of
galaxies. Accordingly, we co-added all the selected PCA and HEXTE data 
to form net spectra for analysis.  Spectra from the individual PCA
detectors were also coadded, and the small differences in gain were
accounted for in the generation of the energy response matrices, 
following procedures prescribed by the PCA analysis team.

A small energy dependence for systematic errors (e.g., Gruber et al 2001, 
Wilms et al. 1999) averaging about 0.3\%, was applied to the PCA 
spectrum. Additionally, PCA spectral channels below 4\,keV and above 
22\,keV were excluded because of sensitivity to artifacts in the 
background model, as well as the rapidly declining effective area of 
the PCA outside these bounds. The HEXTE data were restricted to the 
energy range 19--80\,keV for similar reasons, resulting in source and 
background counting rates of 0.75 and 84.1 count/s respectively 
(cluster~A), and 0.49 and 58.1 count/s (cluster~B). 

Repeating the procedure in RGB, we fit the joint PCA and HEXTE spectra
summed over all observations to three simple spectral models: a 
Raymond-Smith (R-S) thermal plasma emission model, R-S plus a power law 
model, and two R-S models at different temperatures. In all three cases 
most of the observed flux is in a primary $\sim\, 8$ keV R-S component. 
Best-fit parameters and 90\% confidence intervals are listed in Table 
1. The best-fit temperature in the single isothermal gas model is $7.90 
\pm 0.03$ keV, in good agreement with the range determined from ASCA 
measurements (Honda \ea 1996), and only $\sim 4$\% less than the value 
of 8.2 keV deduced recently from XMM measurements of the {\it central 
10'} region of Coma (Arnaud \ea 2001). This region is much smaller than 
the RXTE $\sim 1^{o}$ field of view. The observed $\simeq 6.7$ keV 
Fe XXV K$_{\alpha}$ line yields an abundance of $0.195 \pm 0.008$ (in 
solar units), quite consistent with previously determined values 
($\geq 0.2$). No cold absorption was measurable, and given the 4 keV 
PCA threshold, none was expected.

The poor quality of the fit to a single isothermal model is apparent 
from the fact that $\chi^2 = 76.8$ for 45 degrees of freedom. Residuals 
have a high-low-high pattern which signals the need for another smooth 
spectral component. When a second thermal component is added, best-fit 
parameters are $kT_1 \simeq 7.5$ keV, and a very high $kT_2 \simeq 37.1$ 
keV, with the second component accounting for a modest fraction, 
$\sim 6\%$ of the total flux. For this fit, $\chi^2 = 48.1$, lower by 
28.8 than the value obtained in the single R-S model. The F-test 
probability of the second component is 0.9992 for two additional 
degrees of freedom. The range of solutions is very large, however, 
because the problem is numerically highly degenerate. If we consider 
the possible sets of ($kT_1$, $kT_2$), the contour $\chi^2 + 4.6$ 
defines the joint 90\% probability contour. The lowest temperature 
combination permitted in this range is $kT_1$ = 5.48 keV and $kT_2$ = 
9.0 keV, with fractions of the 4--20 keV flux, respectively, of 24\% 
and 76\%. The highest temperature combination has $kT_1 \simeq 7.5$ keV 
and $kT_2$ unbounded. Equal 4--20 keV flux contributions are obtained 
with the values $6.5$ and $10.5$ keV for the two temperatures.

A somewhat better fit than the two-temperature model, $\Delta\chi^2$ = 
30.7 (with respect to the single R-S model), is obtained when the second 
component is a power-law, with a photon index of 
$2.1 \pm 0.5$ (90\% confidence). The 4--20 keV flux of the power-law 
component is 8\% of the total. This component comprises most of the flux 
only at energies $\geq 60$ keV. The iron abundance changes negligibly from 
the value obtained with the isothermal fit. The count rate from the combined
thermal and power-law emissions, and that of just the power-law
component, are shown in Figure 1, together with the measurements.

\begin{table*}

\caption{Results of the spectral analysis}

\begin{tabular}{|l|ccc|}
\hline
Parameter        & single R-S        & two R-S       & R-S$+$power-law \\
\hline
$\chi^2/dof$      & 76.8/45        & 48.1/43       & 46.1/43          \\
$kT_1$ (\rm{keV})       &  7.90 $\pm$0.03   & 7.47 $\pm$0.22        & 7.67$\pm$0.1
 \\
\,\, Normalization$^a$  &  0.348$\pm$0.002 & 0.345$\pm$0.020 & 0.334$\pm$0.016
\\
Abundance$^b$      & 0.195$\pm$0.008  & 0.192$\pm$0.010 & 0.202$\pm$0.017   \\
$kT_2$ (\rm{keV})  &                  & $37\pm 29$   &                  \\
\,\, Normalization$^a$  &             & 0.011$\pm$0.011 &            \\
$I_\epsilon$(5\, \rm{keV})
(\rm{$cm^{-2}\,s^{-1}$})    &      &    & (3.3$\pm 1.0)\times 10^{-4}$  \\
Photon index      &                  &                & 2.07$\pm$0.46    \\
\hline
\end{tabular}

\tablenotetext{}{Notes: }

\tablenotetext{}{All quoted errors are at the 90\% confidence level. }

\tablenotetext{a}{e.m. = Raymond-Smith emission measure in units of
10$^{-14} \int N_eN_H dV$ / 4$\pi D^2$, where $D$ is the luminosity 
distance and $N_e$, $N_H$ are the total number of electrons and 
protons, respectively.}

\tablenotetext{b}{Abundance is expressed relative to solar values.}

\end{table*}

\begin{figure}
\centerline{\psfig{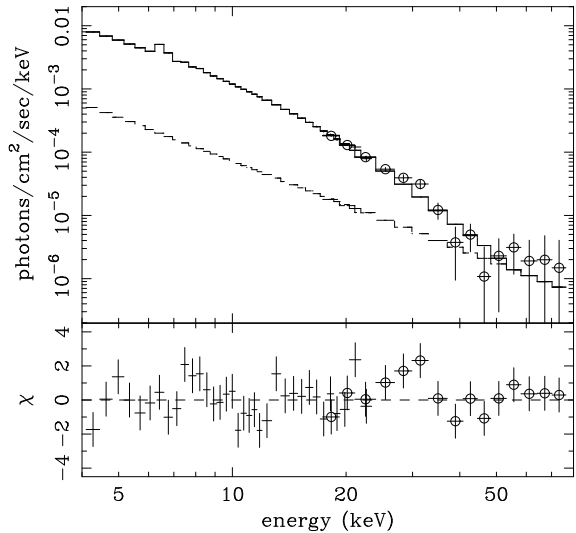}}
\figcaption{The RXTE (photon) spectrum of the Coma cluster and folded 
Raymond-Smith ($kT \simeq 7.67$), and power-law (index $=2.3$) models.
HEXTE data points are marked with circles and 68\% error bars. The 
total fitted spectrum is shown with a histogram, while the lower histogram 
shows the power-law portion of the best fit. The quality of the fit is 
demonstrated in the lower panel, which displays the observed difference 
normalized to the standard error of the data point.}
\end{figure}

Only very limited comparisons can be made with other satellite 
measurements due to differences in spectral bands and field of views.
Using archival ASCA GIS spectra of the central 20' diameter region of
Coma, we find that these are well fit by a single temperature
Raymond-Smith model, with $\chi^2$ = 664 for 713 degrees of freedom.
However, fitting an extra power law component with number index 2.3
reduces $\chi^2$ by 12, and thus is significant at about the 3$\sigma$
level.  The 3-10 keV band common to both ASCA and RXTE is most useful
for comparison: the ASCA power law represents 5 $\pm$ 2 percent of the
total emission, consistent at 68\% with the corresponding RXTE value of
7\%.

EUV emission from the Coma cluster was measured with the EUVE
by Bowyer \& Berghofer (1998). The emission is said to be at a level
significantly higher than what is expected from the main X-ray emitting
gas, and presumably cannot be explained simply by invoking a second,
colder, $T \sim 10^6$ K, gas component. A similar conclusion was recently
reached by Durret \ea (2002) from re-analysis of EUV and ROSAT measurements.

A more direct comparison of our spectral results can be made with 
BeppoSAX observations of Coma with the MECS and PDS experiments. 
Analysis of the PDS measurements led Fusco-Femiano \ea (1999) to 
conclude that a significant (at the $4.6\sigma$ level) NT flux, 
$2.2\times 10^{-12}$ ergs cm$^{-2}$ s$^{-1}$, was detected at energies 
$20-80$ keV. While this range includes the values we inferred from the 
RXTE measurements, it should be noted that an accurate spectral analysis 
of the combined MECS and PDS data is not possible because of the very 
different fields of view of these experiments.

\section{Discussion}

The consistent results from the analysis of the second (year 2000) 
and combined (years 1996 \& 2000) datasets further substantiate the 
reality of the detection of a second component in the Coma spectrum. 
If thermal, this component could help determine the thermal structure 
of IC gas, which in turn would have important implications for the 
use of clusters as cosmological probes in general, and the most 
extensively researched nearby rich cluster in particular. If this 
component is NT emission from a population of \rel electrons, then a 
new dimension for the study of NT IC phenomena will have been opened. 
Since the combined $\sim 264$ ks RXTE observations have not yielded 
direct detection of power-law emission at energies $> 30$ keV, we 
have to invoke other observational and theoretical considerations in 
order to identify more uniquely the origin of the extra emission we 
have deduced.

It is unrealistic to expect that IC gas is fully isothermal outside 
the central region of the cluster. A more likely behavior is at least 
some decrease of the temperature outside a central $\sim 2-3$ core 
radii region. Indeed, previous ASCA measurements of Coma (Honda \ea 1996) 
and other clusters (e.g., Markevitch 1996) seem to have shown some 
deviations from isothermality. In Coma, the Honda \ea (1996) analysis of 
measurements from 14 different pointings of an area extending 
$\sim 1^{o}$ from the center indicated that the IC gas temperature 
varied by $\pm 50\%$ -- with respect to the overall mean value of 
$\sim 8$ keV -- in two azimuthal regions $40'$ from the center. 
However, these two regions cover only a small part of the projected 
area of the cluster, and since the temperature is higher than the mean 
in one region, while it is lower in the other, the overall change of the 
spectral flux, as compared with that from an isothermal gas at the mean 
cluster temperature, is negligible. More recent results from high spectral 
and spatial resolution measurements of Coma with XMM it was concluded 
that the temperature is constant in the central $\sim 10'$ radial 
region, with a best fit value of $8.2 \pm 0.1$ keV (Arnaud \ea 2001). 
While the fits to the XMM data do not seem to have included 
two-temperature models, it is clear that an appreciable emission at a 
significantly different temperature than this mean value would have 
been deduced from the XMM measurements through a larger variance, if 
not in the form of a systematic temperature gradient.

The results of our RXTE analysis yield a statistically most probable 
temperature combination with a prohibitively high value for the second 
component, $kT_2 \simeq 37.1$. At its lowest boundary, the 90\% contour 
region in the ($kT_{1}$, $kT_{2}$) plane does include the 
more acceptable values $kT_{1}=5.5$ keV, and $kT_{2}=9$, but this only 
if the respective 4--20 keV fractional fluxes of these two components 
are 24\% and 76\%. It is quite unlikely that about a quarter of the 
flux could come from a component with a significantly lower temperature 
than the mean value deduced by virtually all previous X-ray satellites. 
In particular, such a component would have been detected in the high 
spatially resolved measurements with ROSAT and XMM.

To assess the possibility that a two-temperature gas model is just a 
simplified representation of a more realistic continuous temperature 
distribution, we have repeated the following simple procedure we 
employed in our analysis of the first RXTE observations (RGB): 
Assuming a polytropic gas temperature profile of the form 
$T(r) \propto n(r)^{\gamma -1}$, with the familiar $\beta$ density 
profile for the gas density, $n(r) \propto (1+r^{2}/r_{c}^2)^{-3\beta 
/2}$, where $r_c$ is the core radius, we calculated the integrated 
flux and the mean emissivity-weighted temperatures as functions of 
$\gamma$, $\beta$, and $r$. These quantities were then calculated in 
the regions $[0, r]$ and $[r, R_0]$ by convolving over the 
triangular response of the PCA with $R_0 \simeq 58'$. From ROSAT 
observations, $r_c \sim 10.0'$, and $\beta \simeq 0.70 \pm 0.05$ 
(Mohr \ea 1999). We sought the range of values of $r$, $\beta$, and 
$\gamma$ for which the two mean emissivity-weighted temperatures 
and respective fluxes from these regions are closest to the values 
deduced from our spectral analysis in Section 3. The results of these 
calculations indicate that for $0.5 \leq \beta \leq 0.9$ and 
$1 \leq \gamma \leq 5/3$, there is no acceptable polytropic 
configuration that matches the observationally deduced values of the 
temperatures and fractional fluxes. For low values of $\gamma$ the 
temperature gradient is too shallow, while for high values the implied 
central temperature is unrealistically high. This simple plausibility 
check suggests that the two thermal components model is somewhat 
inconsistent with the RXTE results. However, the gas distribution 
may be more complicated than considered here, so that a temperature 
structure as implied here cannot be altogether ruled out.

Of particular interest is the somewhat more likely possibility that the 
secondary spectral component is NT. Since emission from an AGN in the FOV 
is not likely (see details in RGB), it is natural to consider that this 
emission is due to Compton scattering of \rel electrons whose 
presence in Coma is directly inferred from many measurements of 
spatially extended region of radio emission (\eg Kim \ea 1990, Giovannini 
\ea 1993). From the measured radio spectral index, $1.34\pm 0.1$, 
it readily follows that the predicted power-law (photon) flux from Compton 
scattering of these electrons by the CMB has an index $2.34 \pm 0.1$, a 
value which is quite consistent with what we have inferred, $2.1 \pm 
0.5$ (all errors are at 90\% confidence). With the measured mean 
radio flux of $0.72 \pm 0.21$ Jy at 1 GHz, the power-law X-ray flux 
deduced here, and the {\it assumption} that the spatial factors in the 
theoretical expressions for the two fluxes are roughly equal, we can 
easily compute (see more details in RGB) the mean volume-averaged 
value of the magnetic field, $B_{rx}$. Taking into account the full 
90\% range of values of the radio flux, radio index, and the power-law 
X-ray flux, we get $B_{rx} \simeq 0.1 - 0.3 \mu$G. This range of 
values for $B_{rx}$ is consistent with our previous estimate (RGB), 
and the range (0.14 - 0.25 \,$\mu$G) deduced by Fusco-Femiano \ea (1999).

Since we have assumed that the spatial factors in the theoretical
expressions for the radio and NT X-ray fluxes are roughly equal, it 
follows that the mean value of the deduced magnetic field is 
independent of the source size and distance. To determine the \rel 
energy density we do have to specify the radius of the emitting region. 
Scaling to the observed radius of the diffuse radio emission, 
$R \sim 20'$, and integrating the electron energy distribution over 
energies in the observed radio and X-ray bands, we obtain $\rho_{e} 
\simeq (8 \pm 3)\times 10^{-14} (R/20')^{-3}$ erg\,cm$^{-3}$; a
distance of $139$ Mpc (with $H_0 = 50$ km\,s$^{-1}$\,Mpc$^{-1}$) was 
used. Based on the high Galactic proton to electron energy density 
ratio of cosmic rays, it can be conjectured that the energetic proton 
energy density is considerably higher than this value.  

The strength of IC magnetic field can also be estimated from Faraday 
rotation measurements of background radio sources seen through clusters, 
yielding a different mean field value, $B_{fr}$. Analyses of such 
measurements usually yield field values that are a few $\mu$G (see, 
\eg, Clarke, Kronberg, and B\"ohringer 2001, and the review by Carilli 
\& Taylor 2002). Clearly, the mean strength of IC fields has direct 
implications on the range of electron energies that are deduced 
from radio measurements, and therefore on the electron (synchrotron 
and Compton) energy loss times. Higher electron energies imply shorter 
energy loss times, with possibly important ramifications for \rel 
electron models (\eg, Rephaeli 1979, Sarazin 1999, Ensslin \ea 1999, 
Brunetti \ea 2001, Petrosian 2001). 

Much has been written about the apparent discrepancy between deduced 
values of $B_{rx}$ and $B_{fr}$. Indeed, it is sometimes claimed that 
this discrepancy makes Compton interpretation of cluster power-law 
X-ray emission untenable. However, $B_{rx}$ and $B_{fr}$ are actually 
quite different measures of the field: Whereas the former is essentially 
a volume average of the \rel electron density and (roughly) the square 
of the field, the latter is an average of the product of the line of 
sight component of the field and gas density. All these quantities vary 
considerably across the cluster; in addition, the field is very 
likely tangled, with a wide range of coherence scales which can 
only be roughly estimated. These make the determination of the field by 
both methods considerably uncertain. Thus, the unsatisfactory observational 
status (stemming mainly from lack of spatial information) and the intrinsic
difference between $B_{rx}$ and $B_{fr}$, make it clear that these two
measures of the field cannot be simply compared. Even ignoring the large
observational and systematic uncertainties, the different spatial
dependences of the fields, \rel electron density, and thermal electron
density, already imply that $B_{rx}$ and $B_{fr}$ will in general be
quite different. This was specifically shown by Goldshmidt \& Rephaeli
(1993) in the context of reasonable assumptions for the field morphology, 
and the known range of IC gas density profiles. It was found that $B_{rx}$
is typically expected to be smaller than $B_{fr}$. Various statistical 
and physical uncertainties in the Faraday rotation measurements, and their 
impact on deduced values of IC fields, were investigated recently by 
Newman, Newman \& Rephaeli (2002); their findings further strengthen the 
assessment that a simple minded comparison of values of $B_{rx}$ and 
$B_{fr}$ is meaningless, and that it is quite premature to draw definite 
conclusions from the apparent discrepancy between values deduced by 
these very different methods to measure IC magnetic fields.

As we have mentioned in the previous section, the excess EUV emission in
Coma could also be NT, and based on the similar morphologies of the EUV
emission and low frequency radio emission, Bowyer \& Berghofer (1998)
interpreted this emission as Compton scattering of the CMB by a population
of low energy electrons. They adopted a value for the power-law index which
is somewhat lower than the value used here, but deduced a similar value
($\sim 0.2\, \mu$G) for the mean magnetic field. More recently, Tsay, Hwang
\& Bowyer (2002) have explored whether a Compton origin for the observed 
EUV excess can be maintained even if the field is as high (few $\mu$G) as 
is currently deduced from Faraday rotation measurements. They conclude that 
this is possible within a limited class of lower energy ($\sim 100$ MeV) NT 
electron models, and that in this case a different (second) population 
of \rel electrons is required to explain the measurements of NT X-ray 
emission by RXTE and BeppoSAX. Note that significant IC density of 
sub-relativistic electrons could in principle also produce high energy 
X-ray emission by NT bremsstrahlung (Kaastra \ea 1998, Sarazin \& Kempner 
2000). However, the properly normalized contribution of such electrons to 
the power-law emission deduced here from the RXTE measurements is too 
small (Shimon \& Rephaeli 2002) to affect our estimated value of the 
magnetic field.

Further evidence for the NT nature of the second spectral component in 
Coma could possibly come from the scheduled 500 ks observation of this 
cluster with with IBIS imager aboard the INTEGRAL satellite. The 
moderate $\sim 12'$ spatial resolution of IBIS can potentially 
yield crucial information about the location and size of high energy 
NT X-ray emission.

\acknowledgments
We are grateful to the referee for useful comments made on an earlier 
version of the paper. 

\parskip=0.02in
\def\ref{\par\noindent\hangindent 20pt}


\noindent

\begin{references}
\ref{Arnaud, M., \ea 2001, A\&A, 365, L67}
\ref{Bowyer, S., \& Berghofer, T.W. 1998, ApJ, 506, 502}
\ref{Brunetti, G., \ea 2001, MN, 320, 365}
\ref{Carilli, C.L., \& Taylor, G.B. 2002, ARAA, in press (astro-ph/0110655)}
\ref{Clarke, T.E., Kronberg,  P.P., \& B\"ohringer, H. 2001, ApJ, 547, L111}
\ref{Donnelly, R.H, \ea 1999, ApJ, 513, 690}
\ref{Durret, F., \ea 2002, astro-ph/0204345}
\ref{Ensslin T.A., \ea 1999, A\&A, 344, 409}
\ref{Fusco-Femiano, R., \ea 1999, ApJ, 513, L21}
\ref{Fusco-Femiano, R. \ea 2000, ApJL, 534, L7}
\ref{Giovannini, G., \ea 1993, ApJ, 406, 399}
\ref{Goldshmidt, O., \& Rephaeli, Y., 1993, ApJ, 411, 518}
\ref{Gruber, D.E. \ea 2001, ApJ 562, 499.}
\ref{Gruber, D.E., \& Rephaeli, Y. 2002, ApJ, 565, 877}
\ref{Honda, H., \ea 1996, ApJ, 473, L71}
\ref{Kaastra, J.S. \ea 1998, Nuc. Phys. B, 69, 567}
\ref{Kaastra, J.S. \ea 2000, ApJL, 519, L119}
\ref{Kim, K.T., \ea 1990, ApJ, 355, 29}
\ref{MacDonald, D.R. 2000, unpublished dissertation, U.C. Riverside}
\ref{Markevitch, M. 1996, ApJ, 465, L1}
\ref{Mohr, J.J., Mathiesen, B. , \& Evrard, A.E. 1999, ApJ, 517, 627}
\ref{Newman, W.I., Newman, A.L., \& Rephaeli, Y. 2002, \apj, in press 
(astro-ph/0204451)}
\ref{Petrosian, V. 2001, ApJ, 557, 560}
\ref{Rephaeli, Y. 1977, ApJ, 212, 608}
\ref{Rephaeli, Y. 1979, ApJ, 227, 364}
\ref{Rephaeli, Y. 2001, in `Astrophysical Sources of High Energy
Particles \& Radiation', NATO ASI, edited by Shapiro \ea, Kluwer, p.143}
\ref{Rephaeli, Y., \& Goldshmidt, O. 1992, ApJ, 397, 438}
\ref{Rephaeli, Y., \& Gruber, D.E. 1988, ApJ, 333, 133}
\ref{Rephaeli, Y., Gruber, D.E., \& Blanco, P.R. 1999, ApJ, 511, L21 (RGB)}
\ref{Rephaeli, Y., Ulmer, M., \& Gruber, D.E. 1994, ApJ, 429, 554}
\ref{Sarazin, C.L. 1999, ApJ, 520, 529}
\ref{Sarazin, C.L., \& Kempner, J.C. 2000, ApJ, 533, 73}
\ref{Shimon, M., \& Rephaeli, Y. 2002, ApJ, in press}
\ref{Tsay, M. Y., Hwang, C.Y., \& Bowyer, S. 2002, ApJ, 566, 794}
\ref{Watanabe, M., \ea 1999, ApJ, 527, 80}
\ref{Wilms, J. \ea 1999, ApJ 522, 460}

\end{references}
\end{document}